\documentclass[conference]{llncs}
\usepackage{amssymb,amsfonts,amsmath,subfigure,color}
\usepackage{graphicx,times,latexsym,makeidx,xspace,url,comment,cite,epstopdf}

\newcommand{\bttransp}{\emph{bt.transp\_disposition}}
\newcommand{\bttarget}{\emph{net.utp\_target\_delay}}

\newcommand{\secR}[1]{Section~\ref{sec:#1}}

\newcommand{\secL}[1]{\label{sec:#1}}

\newcommand{\figR}[1]{Fig.~\ref{fig:#1}}
\newcommand{\figLC}[2]{
        \caption{#2}
        \label{fig:#1}
}
\newcommand{\tabR}[1]{Tab.~\ref{tab:#1}}
\newcommand{\tabLC}[2]{
        \caption{#2}
        \label{tab:#1}
}

\begin{document}
\title{ Experimental Assessment of BitTorrent Completion Time in Heterogeneous TCP/uTP swarms }

\author{ Claudio Testa$^1$, Dario Rossi$^1$, Ashwin Rao$^2$, Arnaud Legout$^2$ }
\institute{$^1$ Telecom ParisTech, Paris, France -- \email{first.last@enst.fr}\\$^2$ INRIA Planete, Sophia Antipolis, France -- \email{first.last@inria.fr} }
 
\maketitle

\begin{abstract}
BitTorrent, one of the most widespread used P2P application for file-sharing, recently got rid of TCP by introducing an application-level congestion control protocol named uTP. The aim of this new protocol is to efficiently use the available link capacity, while minimizing its interference with the rest of user traffic (e.g., Web, VoIP and gaming) sharing the same access bottleneck.

In this paper we perform an experimental study of the impact of uTP on the torrent completion time, the metric that better captures the user experience. We run BitTorrent applications in a flash crowd scenario over a dedicated cluster platform, under both homogeneous and heterogeneous swarm population. Experiments show that an all-uTP swarms have shorter torrent download time with respect to all-TCP swarms. Interestingly, at the same time, we observe that even shorter completion times can be achieved under careful mixtures of TCP and uTP traffic. 

\end{abstract}

\section{Introduction}

Though some might argue that network congestion control is a problem that has been studied to death--to which we tend to agree, at least concerning the large amount of literature on the topic--yet the network architecture and usage are undergoing profound changes, that make the study of congestion control issues once more necessary.

As far as the architectures are concerned, recent research has, e.g., addressed the TCP incast problem in data center networks. As far as the usage is concerned, we have lately witnessed to an explosion of new application-layer flow and congestion control algorithms~\cite{youtube11ccrn,bonfiglio09tmm,ledbat_draft}, which are usually implemented at the application-layer over either TCP~\cite{youtube11ccrn} or UDP~\cite{bonfiglio09tmm,ledbat_draft}. Depending on the application they have been built for, these protocols have rather different goals that reflect deep in their design.

In this work, we focus on 
the BitTorrent filesharing protocol that recently replaced TCP by uTP\footnote{Notice that this new protocol has two names: uTP in the BitTorrent community~\cite{bep29}, and LEDBAT in the IETF community~\cite{ledbat_wg}, that we may use interchangeably in the following.}~\cite{bep29,ledbat_wg}, a lower than best effort protocol for data transport on top of UDP. uTP starts from the observation that nowadays the Internet bottleneck is typically at the user ADSL access link: hence, congestion typically happens between different flows of the same user. Moreover, since ADSL modems have rather long buffers (up to few seconds~\cite{itc22nec,pam10}), using TCP for non-interactive but massive data downloads has a possibly negative impact on interactive communication (e.g., Skype, gaming, etc.). Indeed, as TCP fills the buffer, this self-induced congestion translates into high latency and losses, that possibly harms other interactive application. To avoid the troubles of self-induced congestion and at the same time be efficient for massive data download, uTP tries to limit the end-to-end delay (by reaching a fixed amount of \emph{target delay} in the access buffer) while maximizing the utilization of the access capacity at the same time. As the queuing delay is bounded with uTP, this improves the user experience for interactive applications.

While uTP has sound and appealing goals, it is clearly understood that users will be the ultimate judge of BitTorrent performance, as in BitTorrent's own words \emph{``unless we can offer the same performance [of TCP], then people will switch to a different BitTorrent client"}~\cite{bt_utp}. Our recent work~\cite{p2p11} suggests that uTP performs better than standard TCP, as the use of uTP practically limits the queuing delay to a small target, this translates into faster signaling as a side effect. However, results in~\cite{p2p11} are based on ns2 simulation: it becomes thus imperative to assess whether the observed phenomena also happens in practice, which is precisely the scope of this work.

We run BitTorrent applications in a flash crowd scenario over the Grid'5000 platform~\cite{grid_site}, with special attention to the main user-centric metric, the torrent completion time.  Results of our experiments confirm our previous simulation results, in that, as observed in~\cite{p2p11}, uTP can reduce the torrent download time. 

Yet, this experimental campaign brings new insights beyond~\cite{p2p11}. 
Currently, the default settings BitTorrent yield to the use of a \emph{mixture of TCP and uTP traffic}: as such, in this work we evaluate how this choice performs compared to the cases in which all peers use only a single protocol between TCP and uTP. In this case, our experimental results show that completion time under heterogeneous swarms can be even lower than all uTP (and, clearly, all TCP) swarms.

The remainder of this paper is organized as follows. First, \secR{testbed} reports preliminary insights on low-level BitTorrent settings gathered from a small local testbed, which are instrumental to our experiments, whose results are reported in \secR{grid5000}. Related work are then reported in \secR{related}, and we summarize the paper and discuss future work in \secR{end}.

\section{Preliminary Insights}\secL{testbed}

As previously stated, the uTP protocol aims at jointly (i) being efficient by fully exploiting the link capacity when no other traffic is present, and (ii) being low priority by yielding to other competing traffic on the same bottleneck. Hence, in order to achieve both these goals, uTP needs to insert only a limited amount of packets in the bottleneck buffer: on the one hand, since the queue is non empty, the capacity is fully exploited. On the other hand, as the queuing delay in the buffer remains bounded, this doesn't harm interactive applications. 

uTP exploits the ongoing data transfer to measure the One-Way Delay (OWD) on the forward path. While measuring the OWD is notoriously tricky among non-synchronized Internet hosts, uTP is interested in the \emph{difference} between the current OWD and the minimum OWD ever observed (used as an approximate reference of the base propagation delay).  In turn, this OWD difference yield to a measure of the current \emph{queuing delay}, that is used to drive the congestion window dynamics: when the measured queuing delay is below a given \emph{target delay}, the congestion window grows, but when the queuing delay exceeds the target the congestion window shrinks.

The impact of this new protocol on the performance of BitTorrent can be affected by essentially two different settings. At a single flow level, uTP is primarily driven by the uTP \emph{target delay} setting. At a swarm level, peers \emph{relative  preference} for TCP vs UTP protocols likely plays an important role as well. Hence, before running a full-fledged set of experiments, we need to get some preliminary insights on the settings of the above two parameters. In more details, these are: (i) \bttarget, that tunes the value of uTP delay of each flow, and (ii) \bttransp, that drives TCP vs uTP preference of the BitTorrent client.

To this aim, we perform a battery of tests  in a completely controlled environment involving one seed and two leechers, all  running the latest version of the uTorrent client available for Linux (3.0 build 25053, released on March 2011).
Clients in the local testbed are interconnected by a 100\,Mbps LAN and we use \texttt{netem} to emulate  an access bottleneck on the PC running the seed, whose uplink capacity is then capped at 5\,Mbps. The tracker is private within the testbed and is used to announce a set of three different torrent files having different file and chunk sizes (file size of 10, 50 and 100\,MBytes, and chunk size of 256, 512 and 1024\,KBytes respectively).

To understand BitTorrent settings, we tweaked the default configuration\footnote{uTorrent clients store their configuration in a file which is not directly editable (as it also contains an hash value on the configuration content, performed by the client itself) and moreover Linux GUIs do not offer the possibility of modifying the default settings.  However, the configuration file format used by Linux clients is the same as the one of the Windows clients: hence, we used Windows GUIs to produce a pool of configuration files, that we later loaded in Linux.} aiming at (i) verifying the compliance of \bttarget\ to the imposed delay and (ii) understanding how \bttransp\ settings, which controls when uTP is used, impacts the performance of BitTorrent.

In the case of (i) \bttarget, usually a single experiments is sufficient to verify its compliance, since every flow obeys its own settings. Conversely, multiple experiments were necessary in the case of (ii) \bttransp, since the behavior of a peer is affected by the \bttransp\ value of other peers as well. 

In the following we summarize the most relevant findings of the local testbed experiments. Overall, in these tests we captured about 3\,GBytes of packet level traces, that we make available to the scientific community at~\cite{testbed_web}. The remaining of the experiments, performed on the Grid'5000 platform, are reported in \secR{grid5000}.

%
%

\subsection{\bttarget: Target delay settings}
\begin{figure}[t]
\begin{center}
 \includegraphics[angle=-90,width=0.45\textwidth]{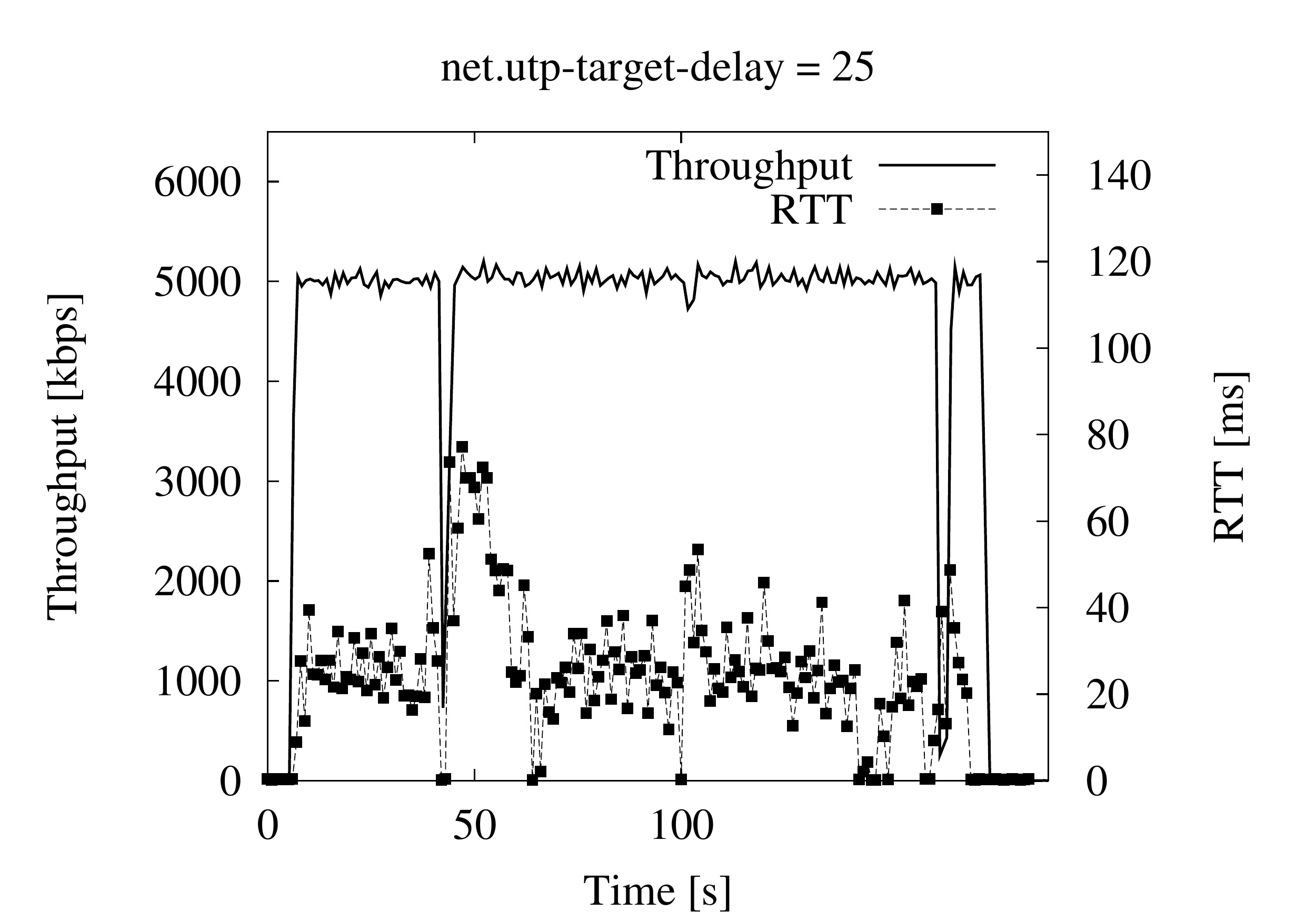}
 \includegraphics[angle=-90,width=0.45\textwidth]{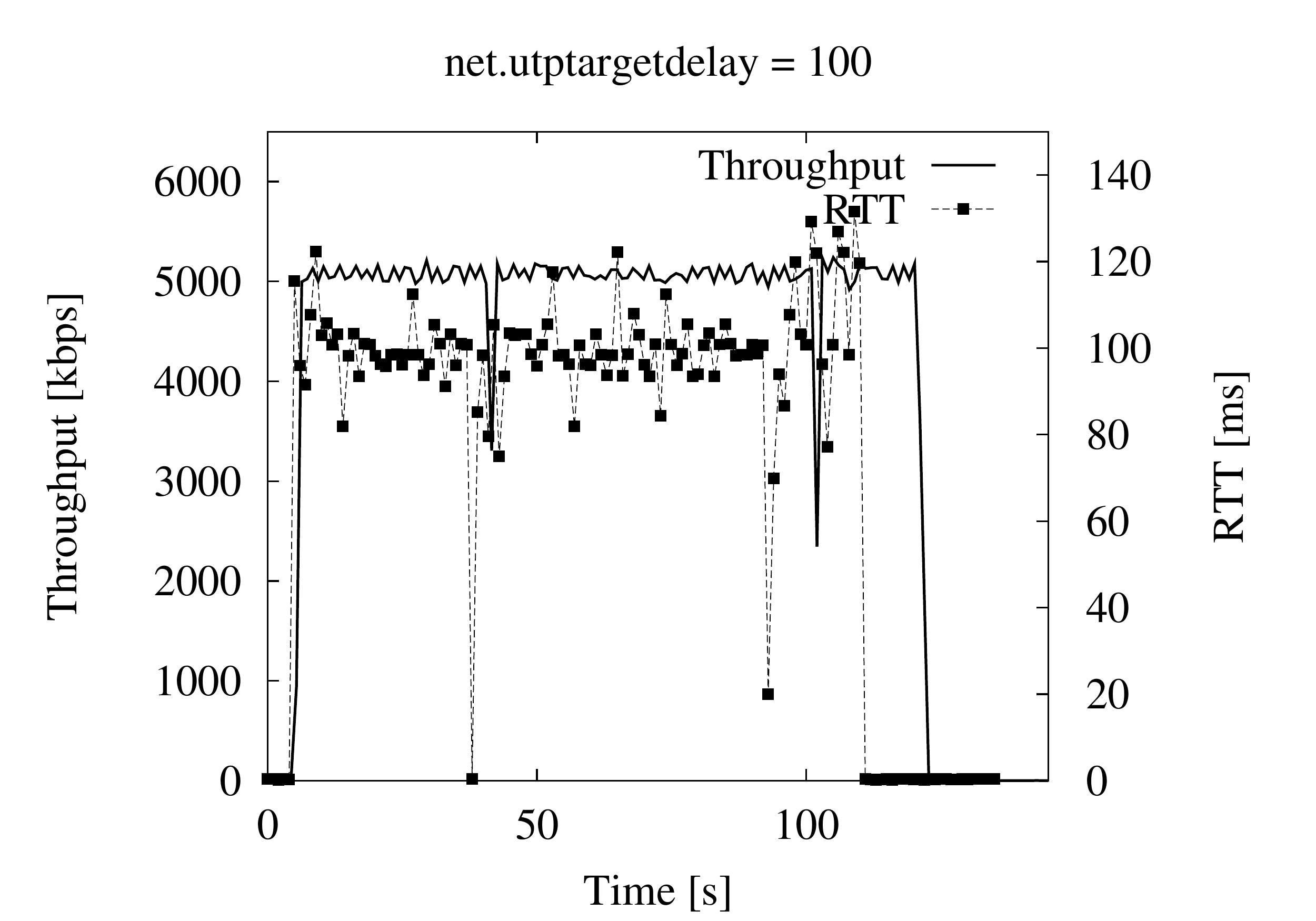}
\end{center}
 \figLC{target}{uTP Target Settings: RTT and throughput for a single flow with  \bttarget=25\,ms (left) and \bttarget=100\,ms (right).}
\end{figure}

The \bttarget\ parameter stores the value of the uTP target delay in milliseconds, and  its default value is equal to 100\,ms as stated in the BEP29~\cite{bep29}. Furthermore, uTP is also standardized at the IETF LEDBAT Working Group~\cite{ledbat_draft}, which specifies 100\,ms as a \emph{ mandatory upper bound value} (while earlier version of the draft referred to a 25\,ms delay target). 

Yet, the GUI of the Windows client allows to modify its default value, opening the way for competition between legacy applications.  This behavior is confirmed by \figR{target}, where we show two experiments, performed at different times, where a single uTP flow sends data on a 5\,Mbps bottleneck, with different values of the uTP target delay. We see in \figR{target} that for both target delay settings, BitTorrent is using the entire available capacity, and the end-to-end delay corresponds to the target that we set by means of the \bttarget\ settings.

As the uTP/LEDBAT specifications~\cite{bep29,ledbat_draft} refer to a mandatory target value, we comply to the standard and focus in the remainder of this paper on the study of swarms with the same default value for \bttarget. At the same time, we point out that as a future work, it would be interesting to see whether, by tweaking the \bttarget\ value, some peers (or some applications) can gather an advantage over the rest of the swarm.  

%


\subsection{\bttransp: TCP vs uTP settings}
\secL{tcpvsutp}

The second parameter, namely \bttransp, controls which protocol is used for the incoming and outgoing data connection of the client. As reported in the online uTorrent manual\footnote{\emph{http://www.bittorrent.com/help/manual/appendixa0212\#bt.transp\_disposition}}, \bttransp\ is a bitmask that sums up the following behaviors: 
\begin{itemize}
    \item 1:  attempt outgoing TCP connections
    \item 2:  attempt outgoing uTP connections
    \item 4:  accept incoming TCP connections
    \item 8:  accept incoming uTP connections
    \item 16: use the new uTP header format 
\end{itemize}
uTorrent default value is 31, which means that the client will accept both TCP and uTP flavors, for either sending or receiving data, possibly using the new uTP header format.

To understand the implications of \bttransp\ settings, we perform a number of tests with heterogeneous settings of the client. Notice that the parameter space we explore and that we make available at~\cite{testbed_web} is larger than the one reported in \tabR{settings}. Yet, for the sake of simplicity, we only report in the table the cases that we later study in \secR{grid5000}, which already conveys some interesting information. Notice also that in all the experiments, the seed is set with the default value \bttransp$=$31.
 
In \emph{Case 1}, the two leechers A and B have different setting for the \bttransp\ parameter: more precisely, A should attempt data connection only using TCP while B should use uTP (and both will accept every flavor in reception). Our experiments show that in this scenario,  peer B sends data to peer A using the uTP protocol, which is the expected behavior. However, peer A sends data to peer B using the uTP protocol too, which happens consistently over all repetitions, and irrespectively of file and chunks size. The reason is that when a uTP connection between B and A is opened by B, A can use this opened bidirectional connection to send data to peer B. Besides, as confirmed by Arvid Norberg, one of the main BitTorrent developers, the uTorrent client has a \emph{hardcoded uTP preference}, so that in case both a TCP and a uTP connection will be successfully established, the former will be closed and only the latter will be used. As we will see in \secR{grid5000}, this will have some important (and beneficial) consequences on the overall swarm completion time.

In \emph{Case 2}, leecher A attempts and accepts only connection via TCP (as a legacy BitTorrent implementation would do), while leecher B maintains the default value for \bttransp\ (which means to attempt and accept both protocols). In this scenario, any communication between the two peers are performed using the TCP protocol, which is consistent and expected for backward compatibility with older BitTorrent client.

Other cases, not shown in \tabR{settings}, yield to different shares of traffic between TCP and uTP. At the same time, since the number of leechers is small, the exact value of the breakdown is heavily influenced by the seeder flavor as well. As such, we defer a quantitative analysis of such a breakdown in the next section.

\begin{table}[t]
    \tabLC{settings}{TCP vs uTP BitTorrent transport disposition.}
    \begin{center}
    \includegraphics[width=0.8\textwidth]{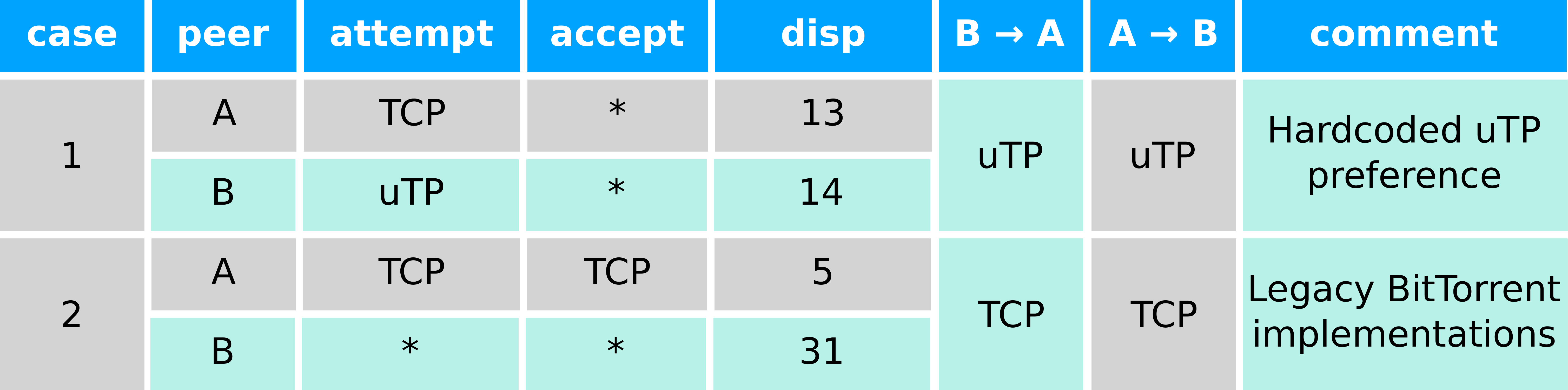}
\end{center}
\end{table}

\section{Experimental Results}\secL{grid5000}

We now report the experimental results on the impact of uTP and TCP transport on the torrent completion time. We first briefly describe the Grid'5000 experimental platform and then focus on two case studies, namely (i) homogeneous and (ii) heterogeneous swarms, depending on the \bttransp\ settings for the leechers.

Homogeneous settings refer to scenarios were all peers have either a TCP-only preference (\bttransp$=$5, which mimic the behavior of old uTorrent versions or legacy applications), or a uTP-only preference (\bttransp$=$10, in case uTP will prevail over TCP), or are able to speak both protocols (\bttransp$=$31, the current default behavior, though with an hardcoded preference for uTP as we have seen in \secR{tcpvsutp}).

Homogeneous settings provide a useful reference, but we must consider also experiments with heterogeneous scenarios that correspond to what is observed in the Internet with clients that do not support uTP at all, or that support uTP but as a fallback choice rather than the default one.

We therefore investigate heterogeneous settings as well, considering scenarios with different ratios of peers using uTP and TCP. More precisely, we consider the case where peers are able to accept any incoming protocol, but have different preferences for the uplink protocol (\bttransp$=$13 for TCP, and \bttransp$=$14 for uTP). We consider the case where preference is split 50/50, 25/75, or 75/25 to mimic scenarios where TCP vs uTP preferences are fairly balanced, or biased toward one of the two protocols. 

Notice that, while there may be several uTP implementations available, different BitTorrent applications use different default settings (i.e., sticking to TCP preference or embracing uTP) depending on the success of the new protocol (and the existence of readily available libraries for different operating systems).

\subsection{Grid'5000 Setup}

We performed experiments on a dedicated cluster of machines that run Linux as the host operating system and using the uTorrent 3.0 client as before.
Hosts of the Grid'5000 platform are interconnected by an high-speed 1\,Gbps LAN, and we emulate realistic bandwidth restrictions and queueing of home gateways by using the \texttt{netem} module for the Linux kernel. As noted in~\cite{ledbat_draft} and experimentally confirmed by~\cite{itc22nec,pam10}, ADSL modems can buffer  up to a few seconds worth of traffic: in our experiments, we set the buffer size $B$ according to the uplink capacity $C$ so that $B/C=1$ second worth of traffic. 

We instrumented the Linux kernel to log the queue size $Q$ in bytes and packets after each dequeue operation, logging also the cumulative number of packets served and dropped by the queue. During our experiments, we disabled the large segment offloading~\cite{mogul03offloading} which ensured that the maximum segment size of the TCP and uTP packets never exceeded the maximum transfer unit (MTU). In each experiment we used the Cubic flavor of TCP, the default for Linux kernels: in reason of our previous work~\cite{pam10}, we may expect Cubic to be more aggressive with respect to the standard TCP NewReno flavor, and more similar to the default TCP Compound flavor adopted in recent versions of Windows.

We use 76 machines on the Grid'5000 platform and consider an Internet flash crowd scenario, where a single seed is initially providing all the content (a 100\,MBytes file) to a number of leechers all arriving at the same time (and never leaving the system). Furthermore, each BitTorrent peer (i) experiences an ADSL access bottleneck~\cite{akella} and (ii) congestion is self-induced by each peer and not by other cross-traffic~\cite{ledbat_draft}. 

 As for (i), we start by considering 3 simple homogeneous capacity scenarios in which we limit the leechers and seed uplink capacity to $C=$1, 2 and 5\,Mbps. For the sake of simplicity, as the qualitative results do not change for different values of $C$, in the following we report the results for $C=1$\,Mbps. While it could be objected that Internet capacity are not homogeneous, we argue that homogeneous scenarios are needed as a first necessary step before more complex and realistic environments are emulated. Additionally, the impact of heterogeneous access capacity is a well known clustering effect~\cite{legoutSIGME07}, that we believe to fall outside of our main aim, i.e., the comparison of TCP vs uTP transport, and that can be studies with a future experimental campaign.

As for (ii), we are forced to map a single peer on each host of the Grid'5000 platform,  as otherwise unwanted mutual influence may take place on multiple peers running on the same hosts. Given the number of hosts $N=76$ we can use, this constrains us on the size of the swarm we investigate. However, we prefer to take a cautious approach, and avoid to introduce the aforementioned mutual influence that could bias in an unpredictable way the results of our experiments (see a discussion on the conclusions). 

We repeat the experiments three times for each settings, to smooth out stochastic variability in the experiments due to BitTorrent random decisions (e.g., chunk selection, choke, optimistic unchoking etc). Also, we make results of our campaign available to the scientific community as for the local testbed at~\cite{testbed_web}.
Overall, the volume of collected data in the Grid'5000 testbed amounts to about 16\,GBytes. Yet, we point out that, in reason of the large number of experiments and seeds, we were not able to store packet-level traces, but only periodic transport-layer (i.e., TCP, UDP traffic amount), application layer (i.e., tracker) and queuing logs.

%
%
\begin{figure}[!t]
    \begin{center}
\includegraphics[angle=-90,width=0.45\textwidth]{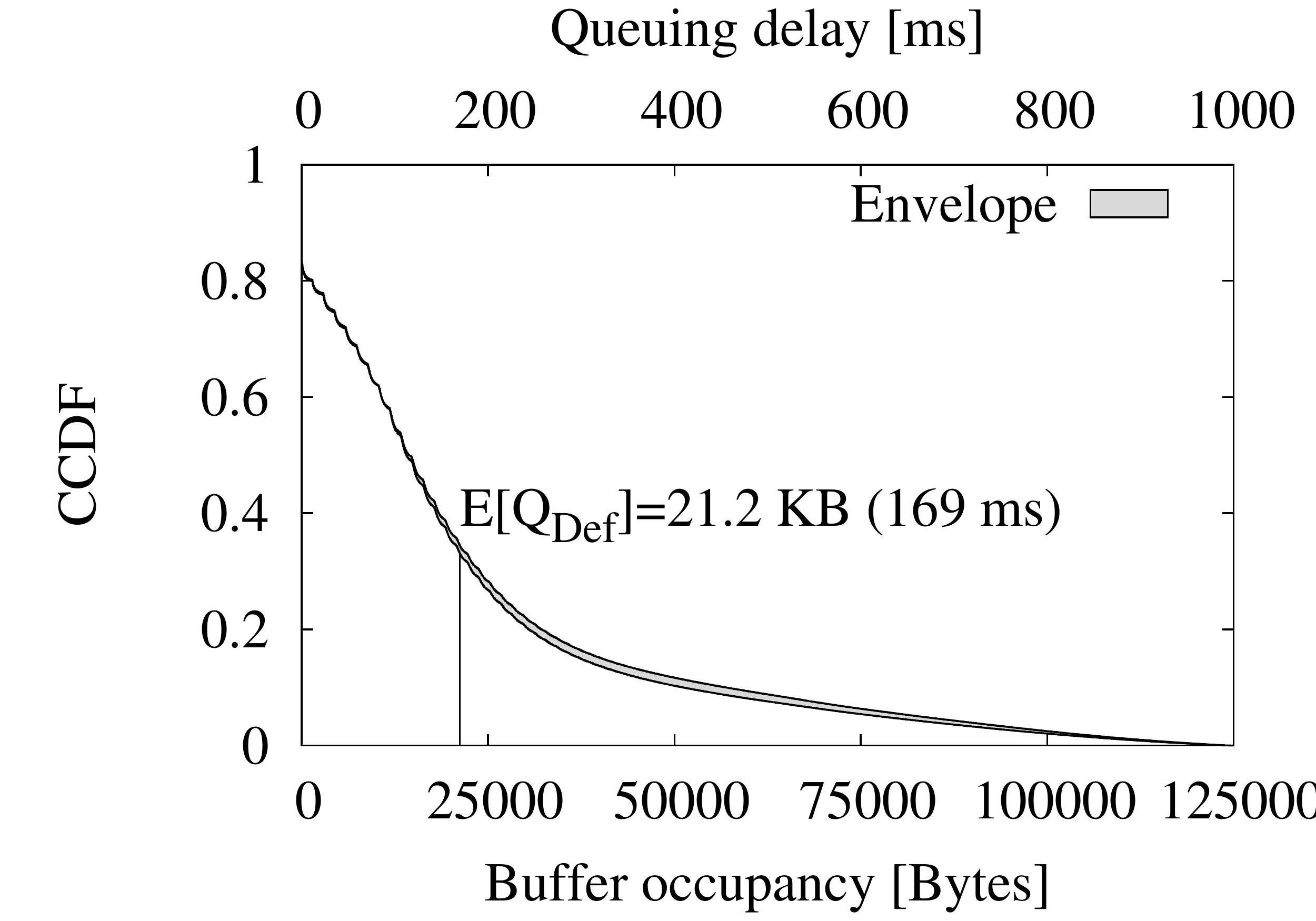}\includegraphics[angle=-90,width=0.45\textwidth]{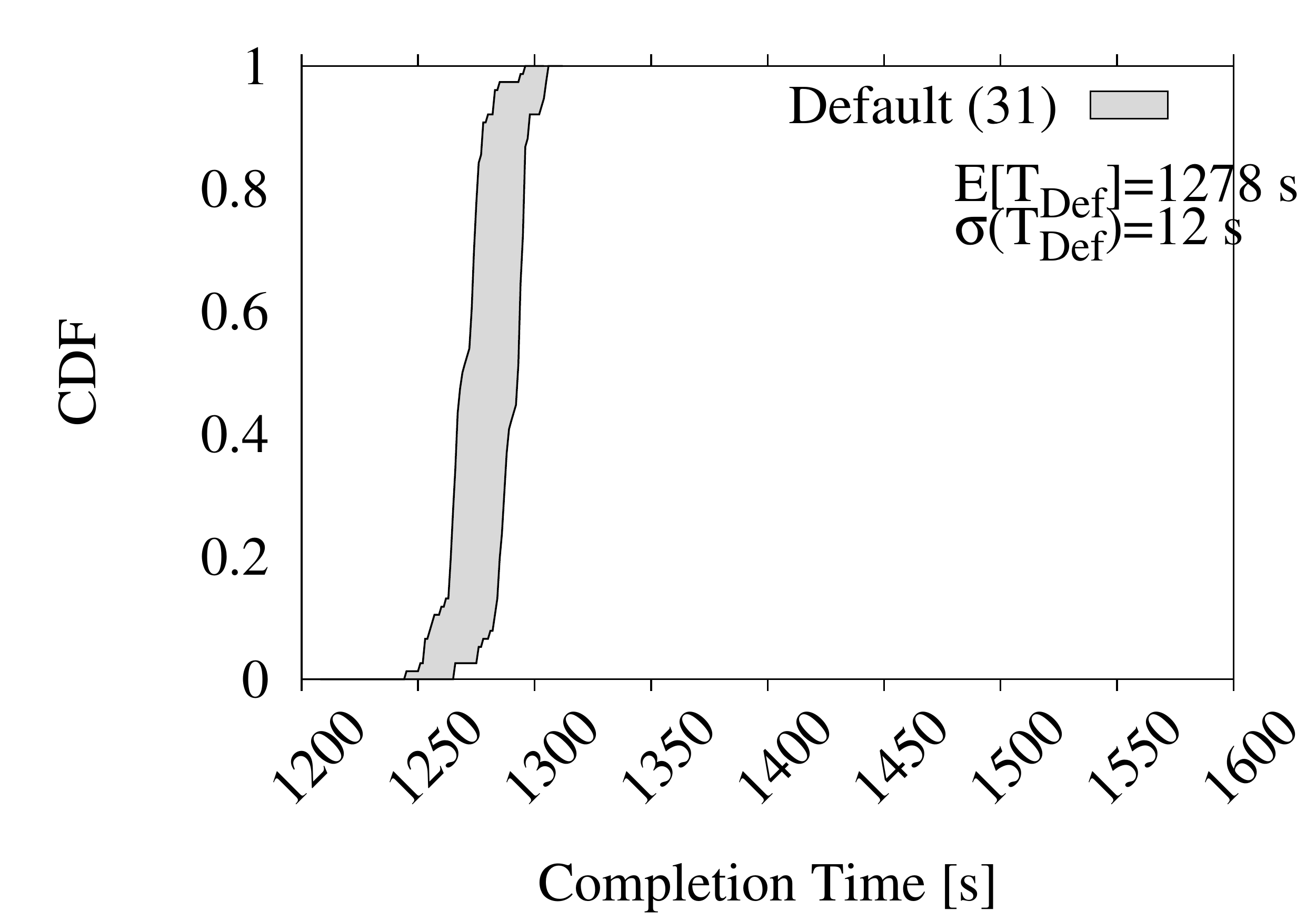} 

\includegraphics[angle=-90,width=0.45\textwidth]{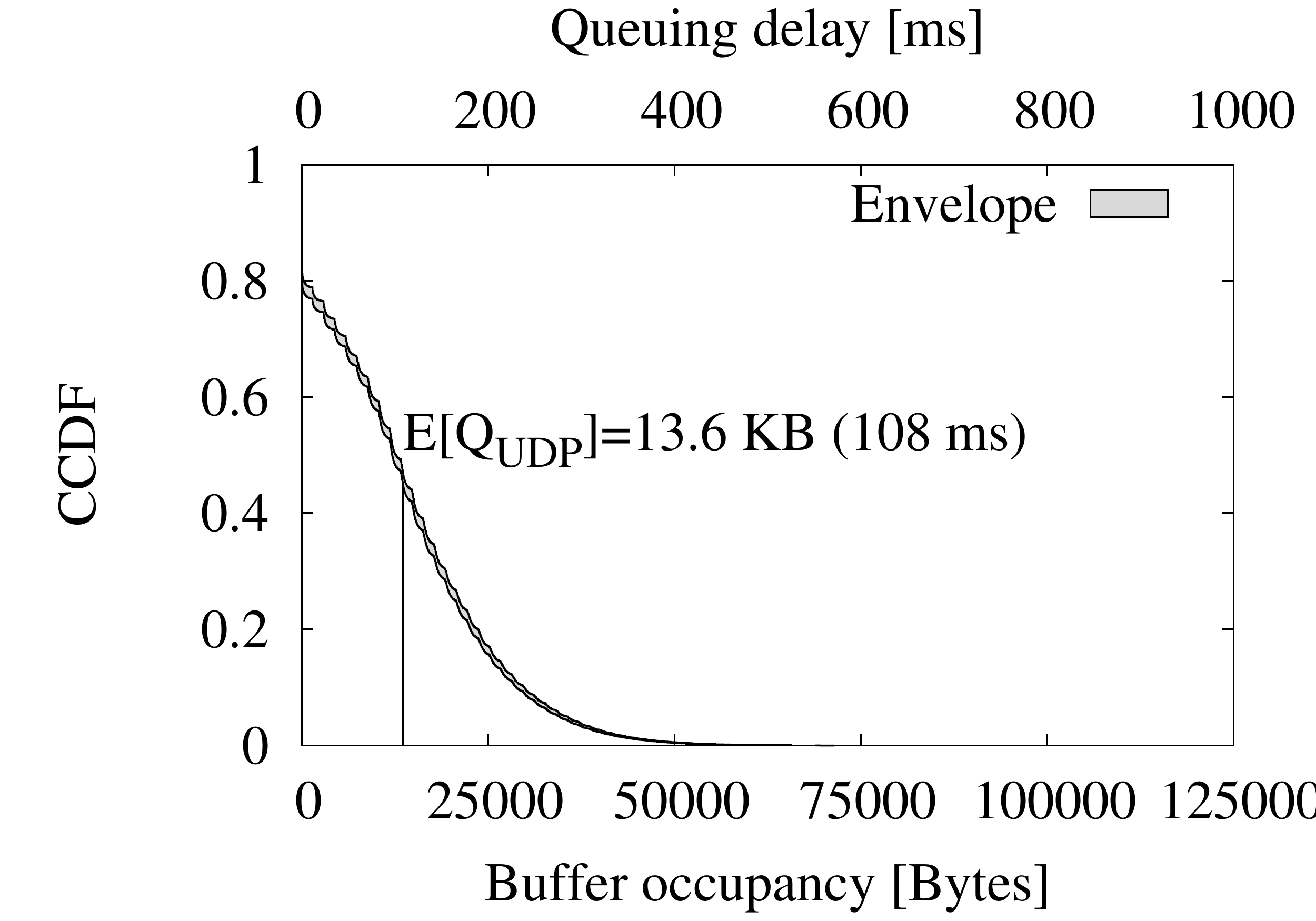}  \includegraphics[angle=-90,width=0.45\textwidth]{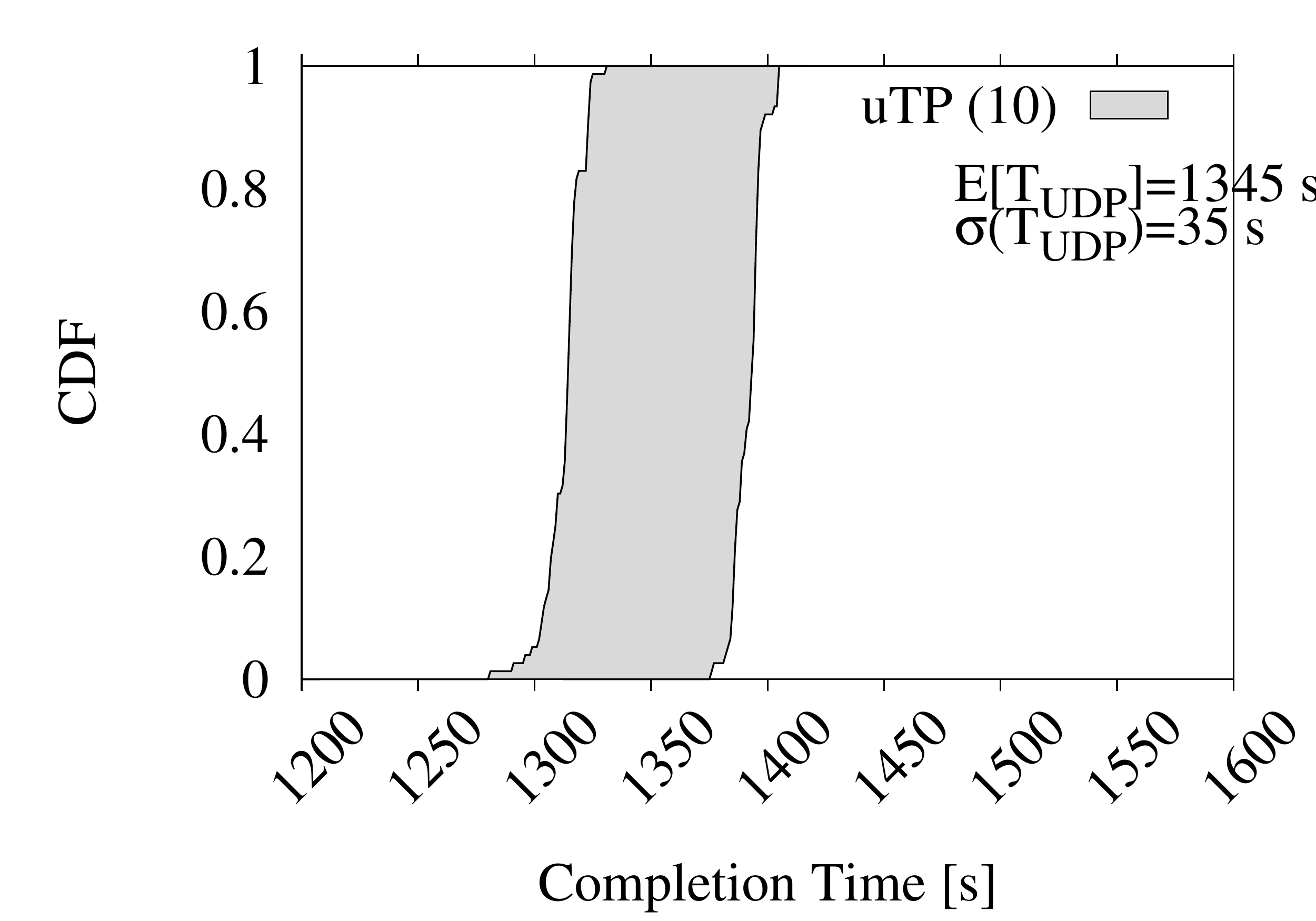}  

\includegraphics[angle=-90,width=0.45\textwidth]{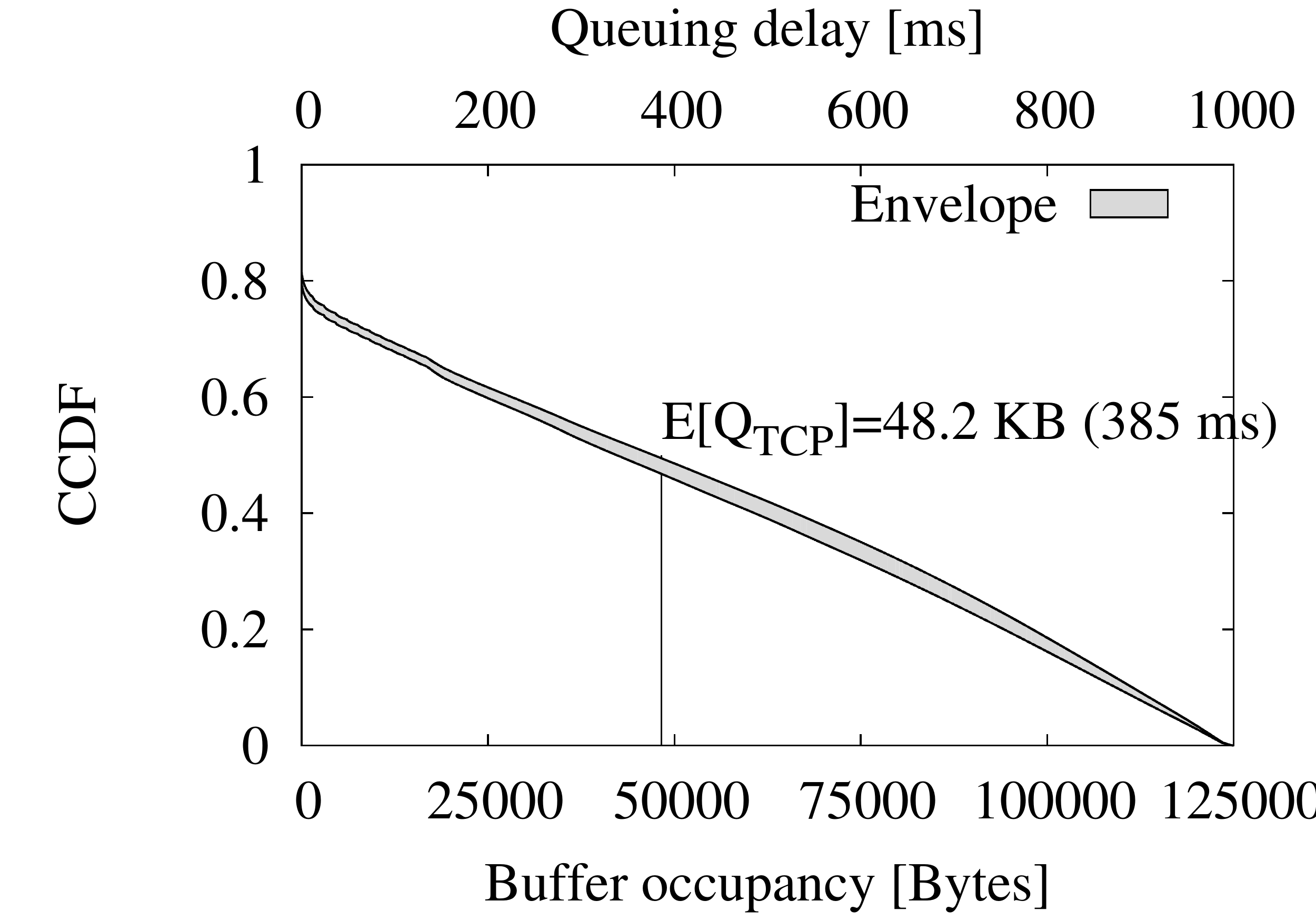} \includegraphics[angle=-90,width=0.45\textwidth]{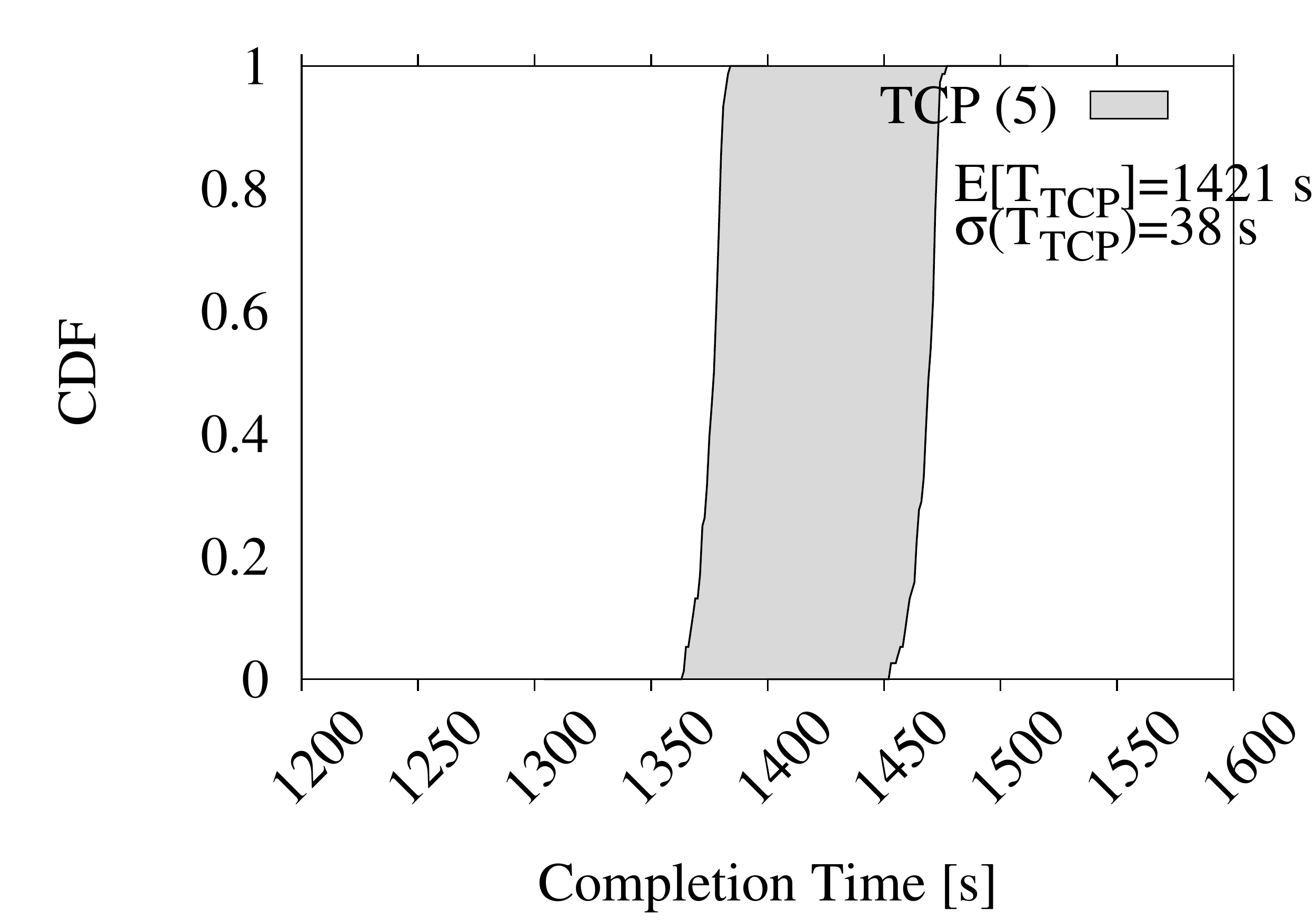}
 	 	  
	      \figLC{homog}{Buffer occupancy CCDF (left) and Completion time CDF (right) for homogeneous swarms: default settings (\bttransp$=$31, top), uTP only swarm (\bttransp$=$10, center) and TCP only swarms (\bttransp$=$5, bottom). The vertical line in the Buffer occupancy plot represent the average of the queue length (in KB and ms).}
    \end{center}
\end{figure}

%

\subsection{Homogeneous \bttransp\ settings}
 
Results of the homogeneous case are reported in \figR{homog}. For each metric of interest, the figure reports the \emph{envelope} of the gathered results, i.e., the minimum and maximum curves over the three iterations.

We express results in terms of (i) the cumulative distribution function (CDF) of the torrent completion time $T$, on the right plots and of (ii) the complementary cumulative distribution function (CCDF) of the buffer size $Q$ of the access link of each peer, on the left plots. The buffer size is expressed both in bytes (bottom x-axis) and in terms of the amount of delay an interactive application would experience for the emulated access capacity (top x-axis).

Additional details are reported in the inset of each figure, showing: (iii) the average $E[T]$ and standard deviation $\sigma[T]$ of the torrent completion time; (iv) the byte-wise share between TCP and uTP, with a notation $X/Y$ that specifies that $X\%$ ($Y\%$) of the bytes are carried over TCP (uTP), with $X+Y=100\%$; (v) the mean queue size $E[Q]$ in KBytes and milliseconds.

In the top row we report the scenario where all peers use default settings (\bttransp$=$31), i.e., the peers are able to speak both TCP and uTP protocols. Middle plots report the case of an all-uTP swarm (\bttransp$=$10), while  all-TCP swarms (\bttransp$=$5) are depicted in the bottom row.

Notice that, on the long run, all swarms achieves similar efficiency: looking at the CDF of the buffer occupancy in \figR{homog} we can see that in roughly 80\% of the time, after a dequeue operation the queue is non-empty. That efficiency is also tied to the BitTorrent system dynamics (e.g., pipelining of the requests, chunk exchange dynamics, etc.). Also the number of packets remaining in the queue after a packet transmission further depends on the congestion control protocol of choice. As expected, TCP AIMD dynamics tend to fill the buffer, while uTP strives (and manages) to limit the queue size.

These behaviors translate into different completion times statistics and, especially, completion times appear to benefit from a mixture of TCP and uTP traffic. We point out that, in the mixed case where BitTorrent peers are able to speak both protocols (\bttransp$=$31), the following happens: two connections, a TCP and an uTP ones are attempted, and in case uTP is successfully opened, it is preferred over the TCP one. This translates into a traffic mixture where about 80\% of the data traffic happens to be carried over uTP.

Notice that the queue size alone cannot explain the difference in the completion time statistics (as otherwise, completion time in all-uTP swarms will be the lowest). Hence, we conjecture this result to be the combination of two effects --on the control and data plane-- that are assisted by the use of uTP and TCP respectively. First, a longer queue size due to  TCP can negatively influence the completion time, by hindering a timely dissemination of \emph{control information} (e.g., chunk interest). The longer the time needed to signal out interests, the longer the time prior to start their download, and their subsequent upload to other peers (which harms all-TCP completion time).

Notice indeed that as in the all-TCP case the one way queuing delay due to TCP may reach up to 400\,ms on average, this entails that RTT for signaling exchanges may be on the order of a second, that can possibly slow down significantly the chunk spreading dynamics. Consider then that BitTorrent is using pipeling to avoid a slowdown of the transfer due to the propagation delay of requests for new chunks. From our experiments, it appears that the pipelining used by BitTorrent is not large enough to deal with delays that might be encountered with xDSL connections.

However, as previously said, the completion times statistics are not fully explained in terms of the queuing delay (as otherwise all-uTP swarms should be the winner). Yet, while uTP limits the queue size, and as such it avoids to interfere with a timely dissemination of control messages, uTP is by design less aggressive than TCP. It follows that TCP may be more efficient for rapidly sharing \emph{chunks in the data plane}. This can in turn harm the all-uTP completion time, that is slightly larger with respect to the default settings \bttransp$=$31.
 
 Interestingly, our previous simulation study~\cite{icccn10} shown that a combination of TCP and uTP can increase the efficiency on the case of two flows sharing a bottleneck link. Shortly, this happens because the low-priority protocol is still able to exploit the capacity unused by TCP (as its rate increases when queuing is low), without at the same time increasing the average system queueing delay (as its rate slow down when TCP one increases).
 The experimental results of this work further confirm that a combination of TCP and uTP can be beneficial to the completion time of the whole swarm as well. Moreover, although the exact shape of the completion time CDFs differ across experiments (due to the stochastic nature of BitTorrent chunk scheduling and peer selection decisions), the results are consistent across all iterations.


Unfortunately, latest versions of uTorrent do not allow to export chunk level logs, which could bring further information as the trading dynamics between peers, that remains an interesting direction for future work. 

\subsection{Heterogeneous \bttransp\ settings}

%
%
\begin{figure}[!t]
    \begin{center}
\includegraphics[angle=-90,width=0.45\textwidth]{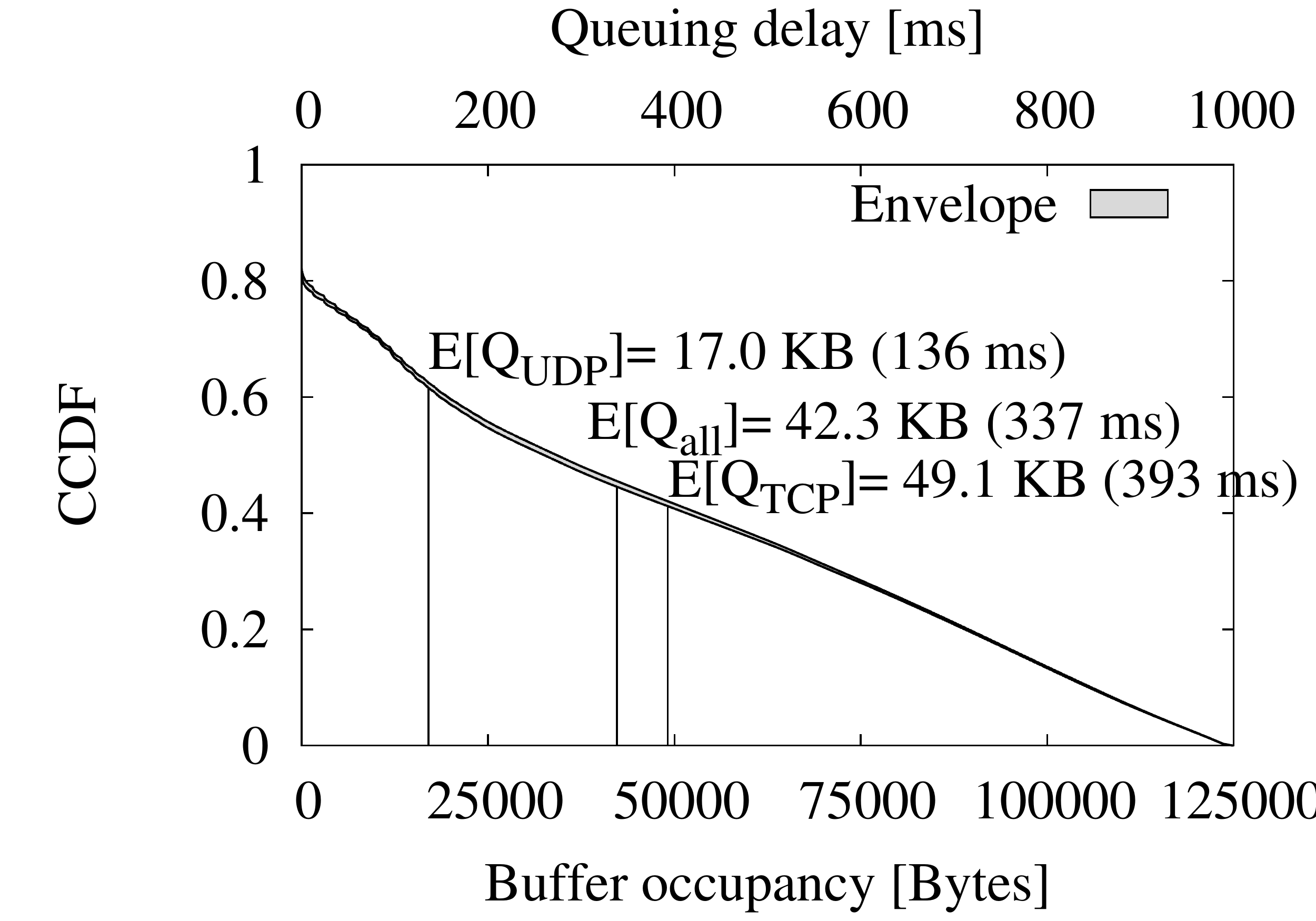}\includegraphics[angle=-90,width=0.45\textwidth]{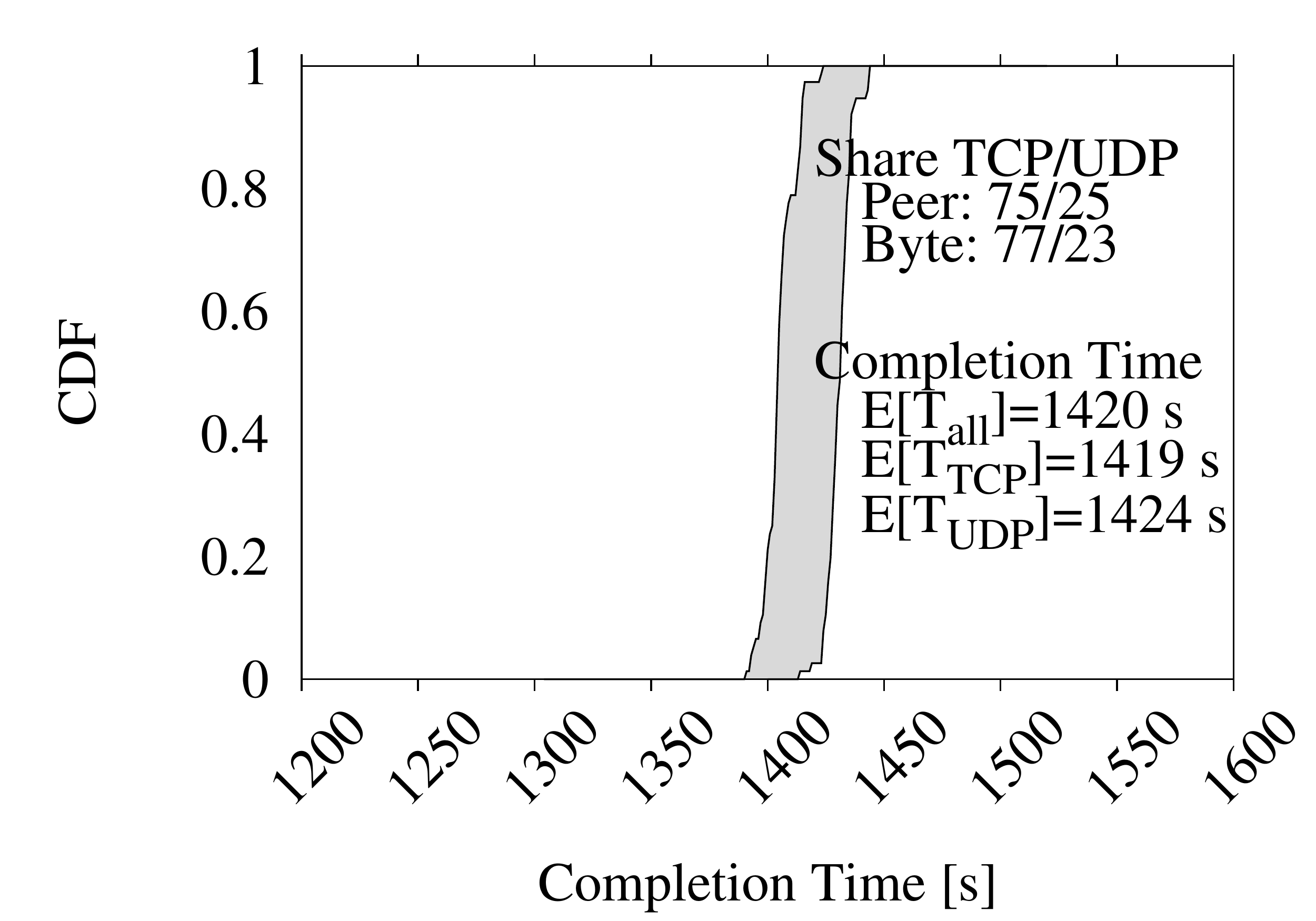} 

\includegraphics[angle=-90,width=0.45\textwidth]{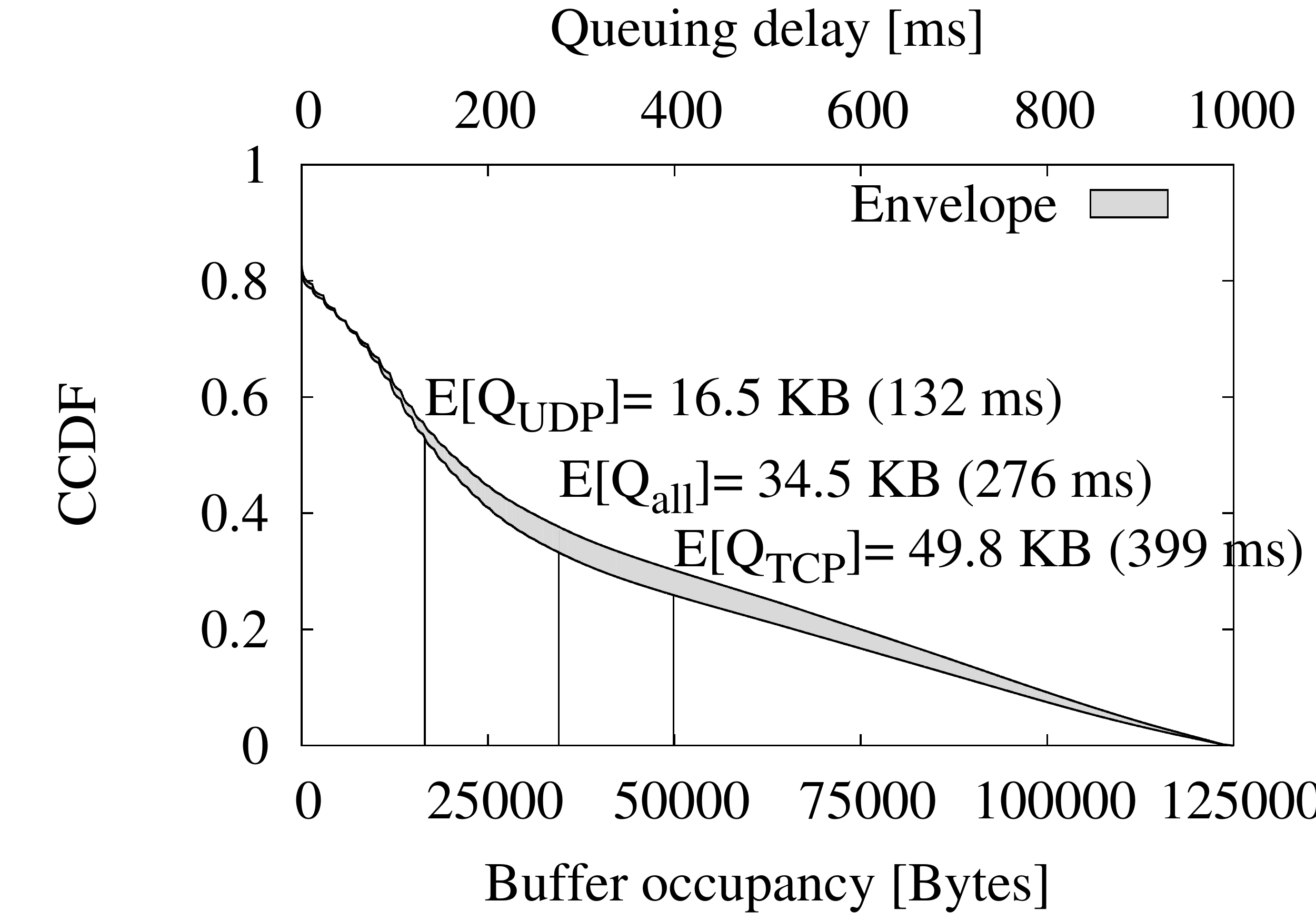}  \includegraphics[angle=-90,width=0.45\textwidth]{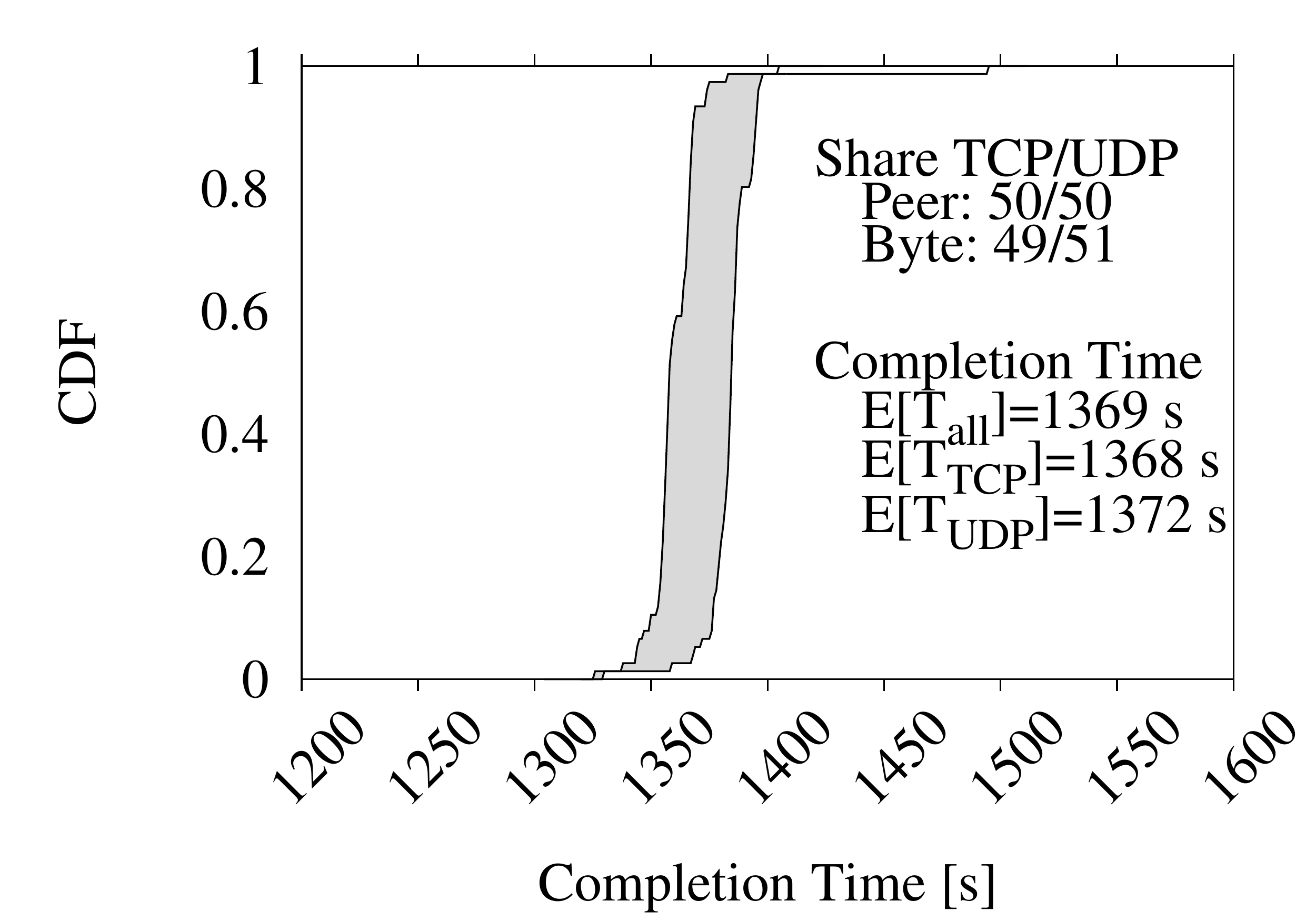}  

\includegraphics[angle=-90,width=0.45\textwidth]{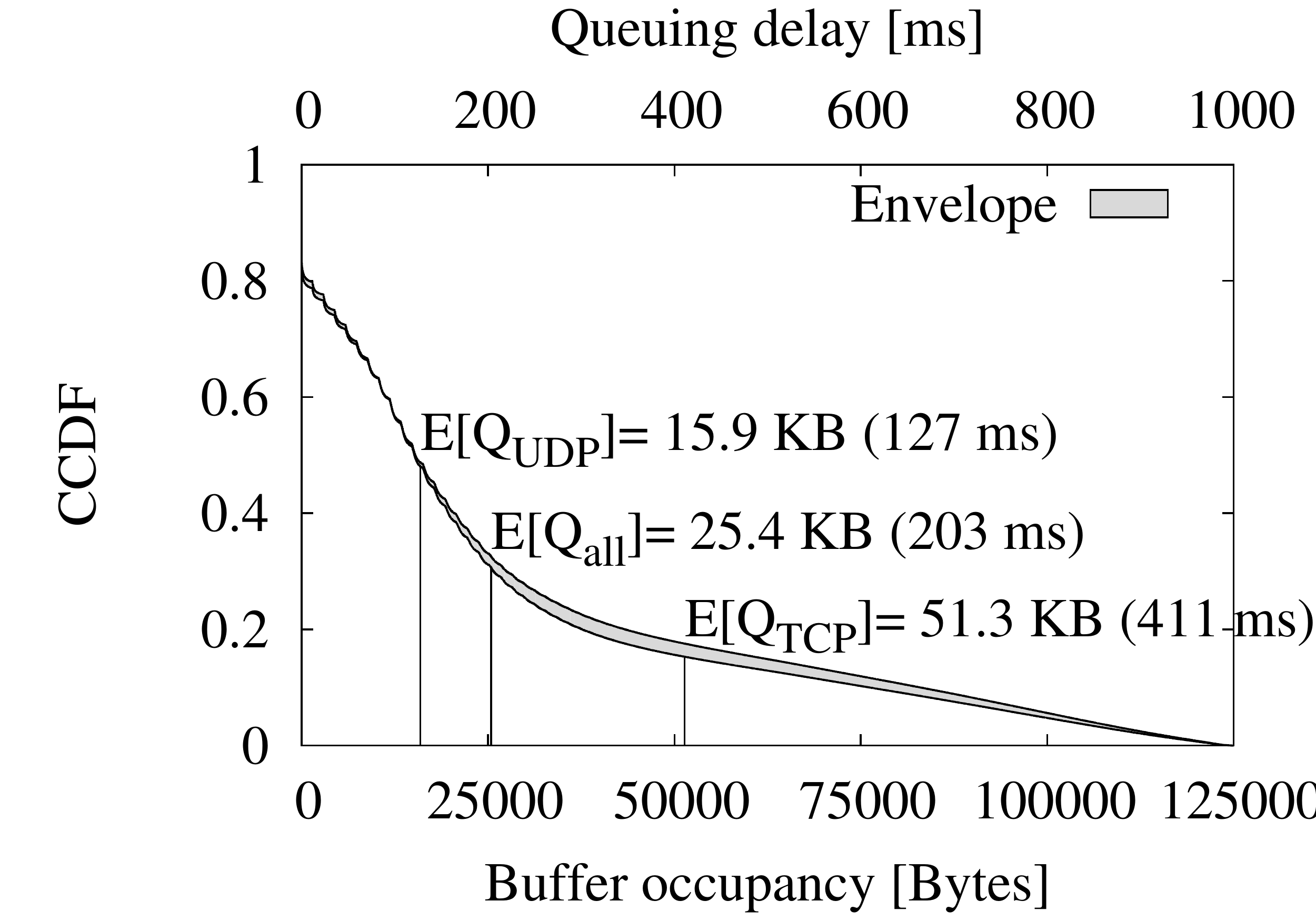} \includegraphics[angle=-90,width=0.45\textwidth]{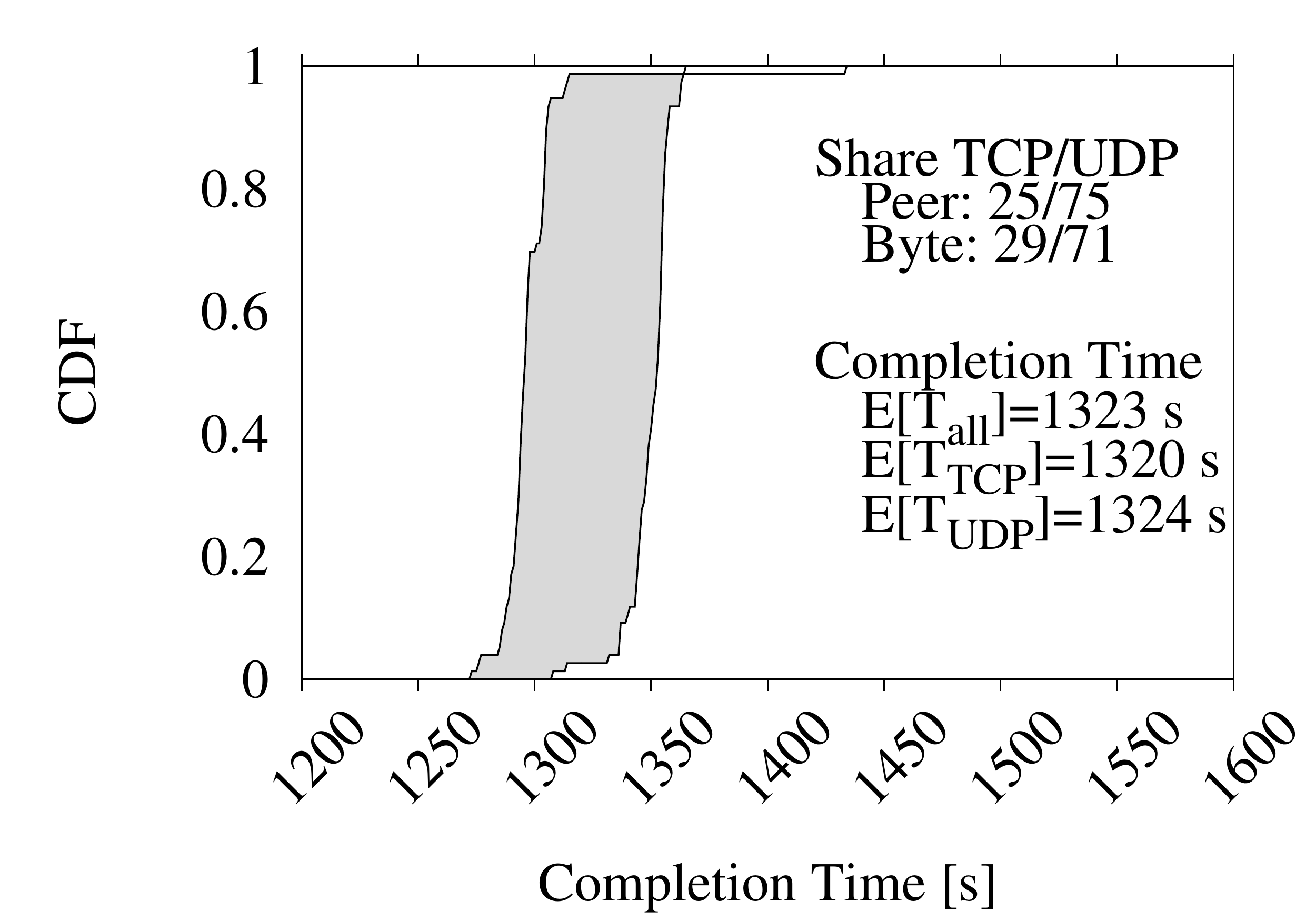}
 	 	  
	      \figLC{heter}{Buffer occupancy CCDF (left) and Completion time CDF (right)  for heterogeneous swarms: prevalence of TCP peers (75/25, top), fair population share (50/50 middle), and prevalence of uTP peers (25/75, bottom).}
    \end{center}
\end{figure}

Having seen that a mixture of TCP and uTP protocols can be beneficial to the completion time, we further investigate different shares of TCP (\bttransp$=$13) vs uTP peers (\bttransp$=$14), i.e., peers that prefer one of the two protocols for active connection open, but that can otherwise accept any incoming connections.

We consider three peer-wise shares, namely 25/75, 50/50, and 75/25 (in the $X/Y$ notation, $X\%$ represents the percentage of peers preferring TCP on their uplink, i.e., \bttransp$=$13). These shares represent three different popularity cases of uTP, that can be either the default in only few implementations of the BitTorrent applications (25/75), or compete equally (50/50) or even be dominant (75/25). We believe these shares to represent illustrative points, covering all relevant scenarios of the possible population repartition. 

The plots in~\figR{heter} additionally report (i) the average queuing delay, for all  the swarm as well as for different peer classes, (ii) the peer- and  byte-wise traffic shares, and (iii) the average system completion time, as well as the average completion time for peers of different classes.

As for (i), notice that the average queuing delay statistics are as expected, with an increase of the queuing delay of uTP peers due to bursty acknowledgements in reply to TCP traffic due to TCP peers in the reverse path. As for (ii), notice that the byte-wise share closely follow the peer-wise share. 
Finally, let us focus on (iii) the completion time statistics. Interestingly, as \figR{heter} shows, while a small amount of TCP traffic is beneficial in reducing the overall swarm completion time, a large TCP amount can instead slow down the torrent download for the whole system. 

Further, notice that completion times are practically the same for uTP and TCP peers (with a slight advantage for the latter). Hence, differently from our previous simulation work~\cite{p2p11}, we do not observe an unfairness of completion time between different peer classes within an heterogeneous swarm. This is due to the fact that~\cite{p2p11} considered a simpler model for \bttransp, that neither (i) accounted for TCP peers using an already opened uTP connection in the reverse side nor (ii) for the hardcoded uTP preference.

\subsection{uTP vs TCP in a nutshell}

\begin{figure}[t]
    \begin{center}
\includegraphics[angle=-90,width=0.45\textwidth]{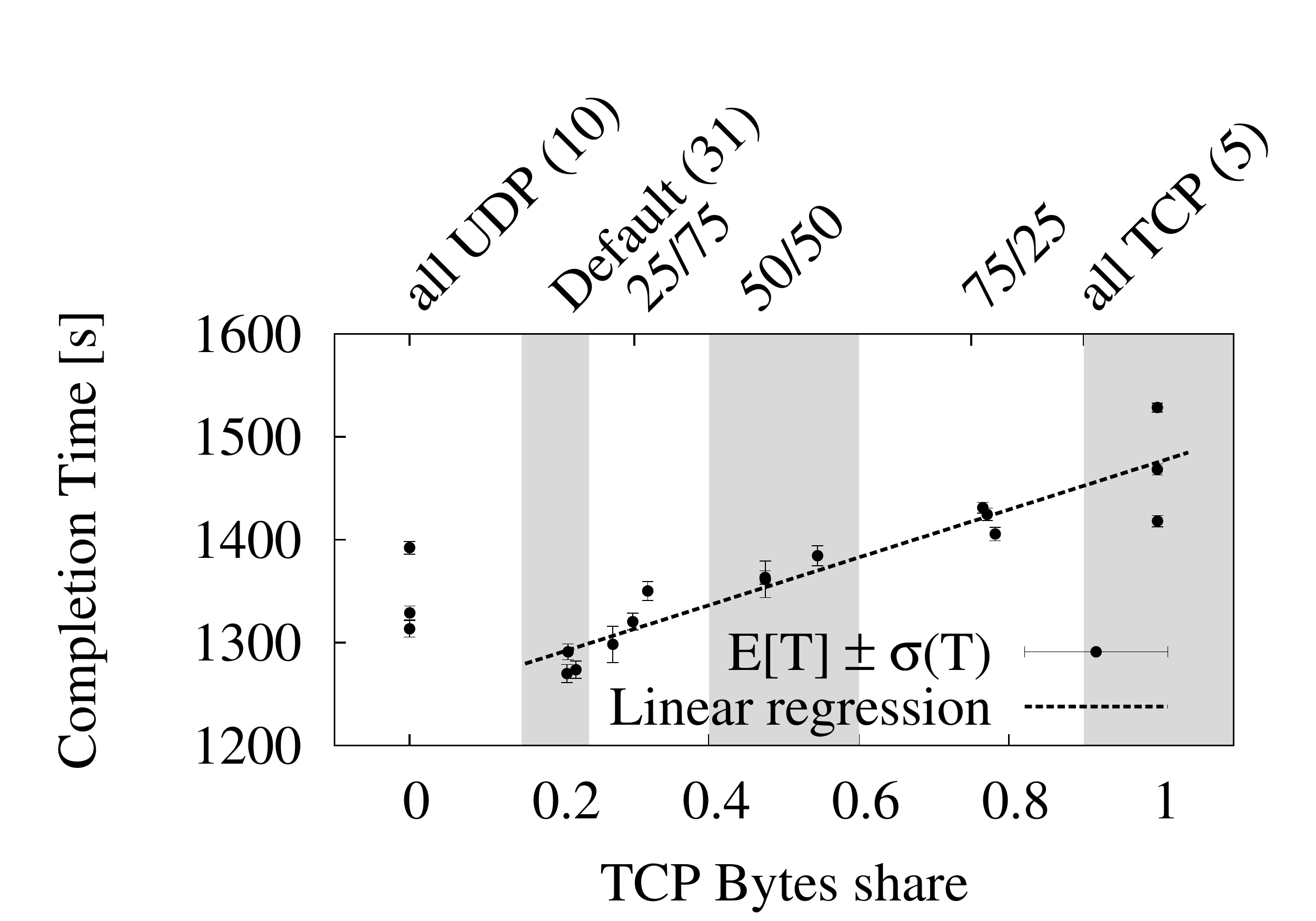}  \includegraphics[angle=-90,width=0.45\textwidth]{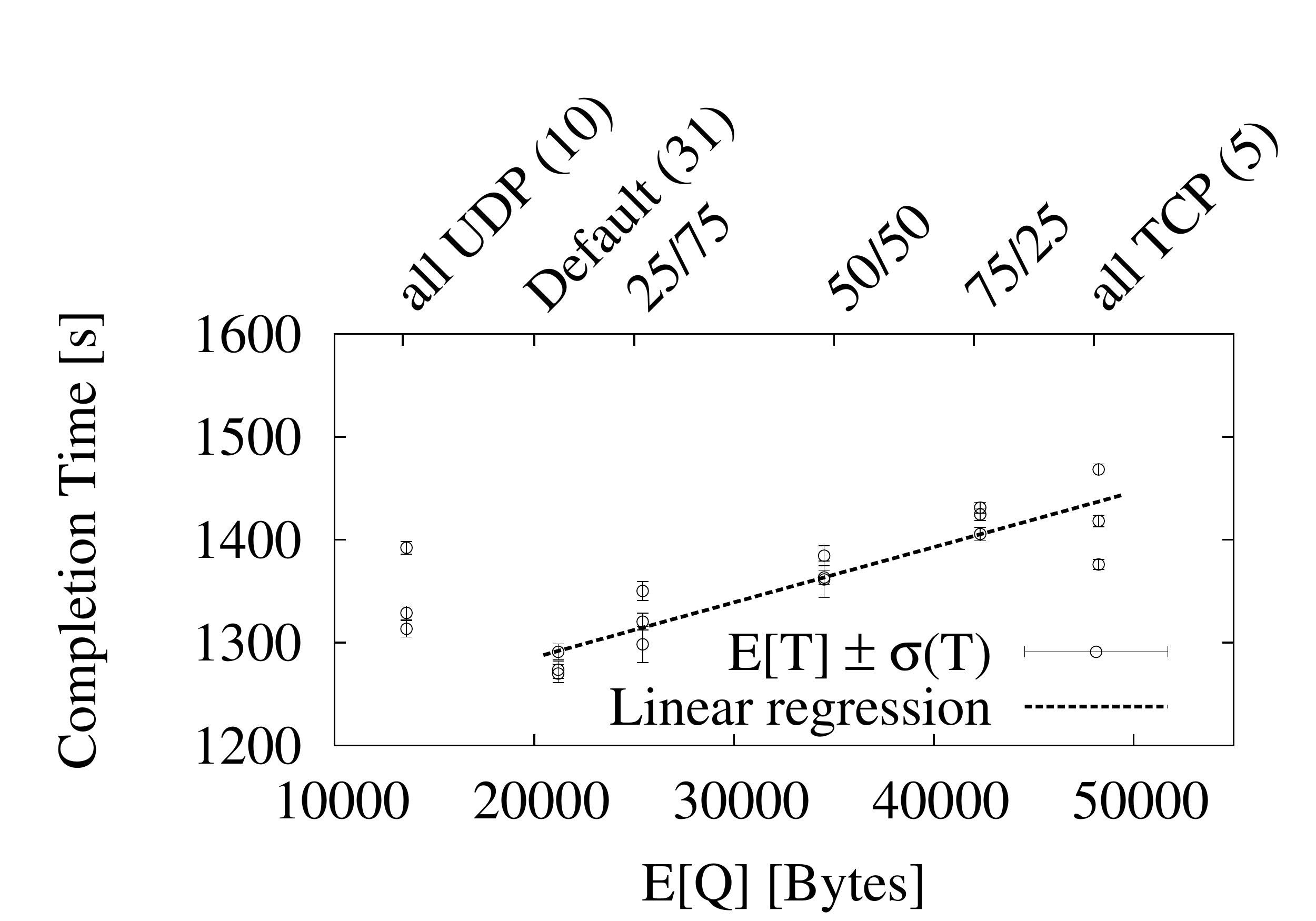} 
	      \figLC{nutshell}{Completion time as a function of the TCP vs uTP byte-wise share (left) and as a function of the average buffer size (right).}
    \end{center}
\end{figure}

\figR{nutshell} present a summary of our results considered so far. $T$ is the completion time (mean and standard deviation) for different iterations with both homogeneous and heterogeneous swarm populations. The $T$ metric is reported as a function of the byte-wise TCP traffic share (left plot) and of the average buffer size (right plot). 

Both plots also report, for TCP traffic shares different from zero (non-0 TCP) only, a linear regression of the completion time. Notice that the linear model provides a nice fit to forecast the completion time performance in presence of different TCP vs uTP mixtures.

Furthermore, as observed in~\cite{p2p11} by means of simulation, \figR{nutshell} confirms that the \emph{completion time increases for increasing buffer sizes,} which in turn generally increases with the amount of TCP traffic exchanged. 

As previously argued, this is due to a slow-down of BitTorrent signaling traffic, while the completion time increase of all-uTP swarms is instead likely due to the low-priority of uTP in the data plane. Hence, we also remark a non-monotonous behavior for the completion time, that decreases for decreasing percentages of TCP traffic, and then increases again for all-uTP swarms. As  different dynamics takes place, hence the linear dependence only applies in case of uTP and TCP traffic mixtures (i.e., non-0 TCP traffic share).

Finally, notice that the default BitTorrent settings consistently yield to the shortest download time we observed in the experiments, which confirms the soundness of the \bttransp\ design decision and settings.

\section{Related work}\secL{related}
Two bodies of work are related to this study: on the one hand, we have work focusing on congestion control aspects~\cite{tcp_nice,tcp_lp,tcp_4cp,youtube11ccrn,bonfiglio09tmm}, and on the other hand work focusing on BitTorrent~\cite{qui04sigcomm,legout10p2p,legoutSIGME07,cohen10iptps,pam10,icccn10,globecom10,lcn10,p2p11}.

First, congestion control literature already proposes several protocols aiming, as LEDBAT, to achieve lower-than-TCP priority, of which TCP-LP~\cite{tcp_lp}, NICE~\cite{tcp_nice}, 4CP~\cite{tcp_4cp} are notable examples. Yet, we pinpoint a recent tendency toward moving congestion and flow control algorithms \emph{from the transport layer to the application layer,} of which uTP~\cite{ledbat_draft} for background file-sharing, Skype~\cite{bonfiglio09tmm} for interactive VoIP and YouTube~\cite{youtube11ccrn} for interactive VoD are again notable examples. 
Unlike transport-layer congestion control, that applies to classes of applications, these application-layer congestion control protocols are usually built for single applications, with specific requirements in mind: these ``one of a kind'' deployments will in our opinion need further attention in the future. Recently LEDBAT also got the attention of Apple's developers, resulting in a implementation for MAC OS, whose preliminary tests are available at~\cite{apple_tests}.

Second, BitTorrent literature dissected many aspects of this successful P2P protocol, from the pioneering time of~\cite{qui04sigcomm}. While our own previous works, such as~\cite{legout10p2p,legoutSIGME07}, already study BitTorrent download performance by means of either passive measurements or experimental tests (as in this work), however they report on performance at a time when BitTorrent was using TCP, and should thus be updated in light of BitTorrent recent evolution. More generally, though related work on uTP exists~\cite{cohen10iptps,cohen10iptps,pam10,icccn10,globecom10,lcn10,gordon2010iccnt,p2p11}, it does however adopt a congestion control perspective (with the exception of~\cite{p2p11}). In particular, an experimental approach is adopted in~\cite{cohen10iptps,pam10,itc22nec}: \cite{cohen10iptps} attacks the problem of clock drift in uTP, while~\cite{pam10} performs a black box study of  initial proprietary versions of the protocol and~\cite{itc22nec} focuses on the interaction of uTP and active queue management techniques that are becoming commonplace in modern home gateways. A simulative approach is instead adopted in~\cite{icccn10,globecom10,lcn10,gordon2010iccnt}: a fairness issue of uTP is revealed in~\cite{icccn10} and solved in~\cite{globecom10}, while~\cite{lcn10} compares the level of low-priority of TCP-LP~\cite{tcp_lp}, NICE~\cite{tcp_nice} and uTP~\cite{ledbat_draft}, and finally~\cite{gordon2010iccnt} investigates policies for dynamic parameter tuning. 

The only previous work addressing the impact of uTP on BitTorrent completion time is our own recent work~\cite{p2p11}, that however employs  ns2 simulations unlike in this work. Interestingly, some of the observations of this study are in agreement with~\cite{p2p11}, e.g., showing a larger completion time for increasing buffer occupancy on the data plane. Yet, we point out that~\cite{p2p11} does not consider an hardcoded preference for uTP, nor bidirectional uTP connections: hence, an interesting difference with respect to the current work is that~\cite{p2p11} forecasted heterogeneous performance for heterogeneous swarms (i.e., larger completion time for TCP peers) that we have shown not to hold on practice.

%
%

\section{Conclusions}\secL{end}

This work assess the impact of uTP (the new BitTorrent congestion control algorithm for data exchange) on the torrent completion time (the main user QoE metrics) by means of an experimental campaign carried on in a fairly large scale controlled testbed.

Our results show that, in flash crowd scenarios, users will generally benefit of a mixture of TCP and uTP traffic, both in homogeneous and heterogeneous swarms.
Interestingly indeed, results with mixed TCP and uTP traffic show consistently shorter download time with respect to the case of homogeneous swarms using either an all-TCP or an all-uTP congestion control.  Especially, our results confirm the soundness of default BitTorrent settings, that yield to the shortest completion time in our experiments.

This results is the combination of two effects, on the control and data plane, that are assisted by the use of uTP and TCP respectively.
By keeping the queue size low, uTP yields to a timely dissemination of signaling information, that would otherwise incur in higher delays due to longer queues building up with TCP. At the same time, by its more aggressive behavior, TCP yields to higher efficiency in the data plane, that results in more timely dissemination of chunk content.

This work leaves a number of interesting points open, that we aim at addressing in the future. First, we would like to investigate whether it would be possible to extend the swarm size by running multiple peers per machine, without however incurring in a bias due to the mutual interaction of the traffic injected. Second, we would like to study the impact of heterogeneous target values in the swarm (i.e., to see whether fairness issues can possibly arise) and refine our experimental setup (e.g., including heterogeneous capacities, peer churn, etc.).  Third, on a longer timescale, we aim at developing a  passive BitTorrent inspector, capable of parsing traffic to produce chunk-level logs, that would greatly enhance our analysis capabilities.

\section*{Acknowledgement}

We thank Arvid Norberg for the fruitful discussions.  Experiments presented in this paper were carried out using the Grid'5000 experimental testbed, being developed under the INRIA ALADDIN development action with support from CNRS, RENATER and several Universities as well as other funding bodies~\cite{grid_site}. Project funding for our future work on the topic would be more than welcome. 

\bibliographystyle{splncs}
\bibliography{biblio}

\end{document}